\documentclass[iop]{emulateapj}
\usepackage{epsfig}

\slugcomment{Submitted to the Astrophysical Journal}

\shorttitle{CXOGBS~J173620.2--293338: A Candidate Symbiotic X-ray Binary}
\shortauthors{Hynes et al.}

\begin{document}

\title{CXOGBS~J173620.2--293338: A Candidate Symbiotic X-ray Binary
  \\Associated with a Bulge Carbon Star}

\author{
Robert I. Hynes\altaffilmark{1,2,3},
M.~A.~P. Torres\altaffilmark{4}, 
C.~O. Heinke\altaffilmark{5},
T.~J. Maccarone\altaffilmark{6},
V.~J. Mikles\altaffilmark{7},
C.~T. Britt\altaffilmark{2,3},\\ 
C. Knigge\altaffilmark{8}, 
S. Greiss\altaffilmark{9}, 
P.~G. Jonker\altaffilmark{4,10,11}, 
D. Steeghs\altaffilmark{9,11},
G. Nelemans\altaffilmark{10},
R.~M. Bandyopadhyay\altaffilmark{12},
C.~B. Johnson\altaffilmark{3}}




\altaffiltext{1}{E-mail: rih@phys.lsu.edu} 
\altaffiltext{2}{Visiting astronomer, Cerro Tololo Inter-AmericanObservatory, National Optical Astronomy Observatories, which are operated by theAssociation of Universities for Research in Astronomy, under contract with theNational Science Foundation.}
\altaffiltext{3}{Department
  of Physics and Astronomy, Louisiana State University, 202 Nicholson
  Hall, Tower Drive, Baton Rouge, LA 70803, USA}
\altaffiltext{4}{SRON, Netherlands Institute for Space Research,
  Sorbonnelaan 2, 3584 CA, Utrecht, The Netherlands}
\altaffiltext{5}{Physics Dept., 4-183 CCIS, Univ. of Alberta, Edmonton AB, T6G 2E1, Canada}
\altaffiltext{6}{Department of Physics, Texas Tech University, Box 41051, Lubbock, TX 79409-1051, USA}
\altaffiltext{7}{I. M. Systems Group, Kensington, MD 20895, USA}
\altaffiltext{8}{School of Physics and Astronomy, University of
Southampton, Southampton SO17 1BJ, United Kingdom}
\altaffiltext{9}{Astronomy and Astrophysics, Department of Physics,
University of Warwick, Coventry, CV4 7AL, United Kingdom}
\altaffiltext{10}{Department of Astrophysics, IMAPP, Radboud University
Nijmegen, Heyendaalseweg 135, 6525 AJ, Nijmegen, The Netherlands}
\altaffiltext{11}{Harvard Smithsonian Center for Astrophysics, 60
Garden Street, Cambridge, MA 02138, U.S.A.}  
\altaffiltext{12}{University of Florida, Gainesville, FL 32611, USA}

\begin{abstract} 
  The Galactic Bulge Survey is a wide but shallow X-ray survey of
  regions above and below the Plane in the Galactic Bulge.  It was
  performed using the Chandra X-ray Observatory's ACIS camera.  The
  survey is primarily designed to find and classify low luminosity
  X-ray binaries.  The combination of the X-ray depth of the survey
  and the accessibility of optical and infrared counterparts makes
  this survey ideally suited to identification of new symbiotic X-ray
  binaries in the Bulge.  We consider the specific case of the X-ray
  source CXOGBS~J173620.2--293338.  It is coincident to within
  1\,arcsec with a very red star, showing a carbon star spectrum and
  irregular variability in the Optical Gravitational Lensing
  Experiment data.  We classify the star as a late C-R type carbon
  star based on its spectral features, photometric properties, and
  variability characteristics, although a low-luminosity C-N type
  cannot be ruled out.  The brightness of the star implies it is
  located in the Bulge, and its photometric properties overall are
  consistent with the Bulge carbon star population.  Given the rarity
  of carbon stars in the Bulge, we estimate the probability of such a
  close chance alignment of any Galactic Bulge Survey source with a
  carbon star to be $\la10^{-3}$ suggesting that this is likely to be
  a real match. If the X-ray source is indeed associated with the
  carbon star, then the X-ray luminosity is around
  $9\times10^{32}$\,erg\,s$^{-1}$.  Its characteristics are consistent
  with a low luminosity symbiotic X-ray binary, or possibly a low
  accretion rate white dwarf symbiotic.

\end{abstract}


\section{Introduction}

In most cool stars, oxygen is more abundant than carbon, resulting in
most carbon being bound in CO molecules and the residual oxygen
forming compounds such as TiO.  Among a minority, the carbon stars,
carbon is more abundant reversing this pattern, and as a result
spectra are dominated by carbon compounds such as C$_2$, CN, and CH.
The carbon stars form a heterogeneous population
\citep{Wallerstein:1998a}.  The modern classification is based on that
of \citet{Keenan:1993a}, and does not fully reflect the likely
evolutionary status of the stars.  Two broad categories of C-R and C-N
stars are identified, based on the earlier R and N spectral types.
C-N stars are the easiest to understand, being asymptotic giant branch
(AGB) stars in which carbon is brought to the surface during the third
dredge-up.  Early C-R stars appear to be core helium burning red-clump
stars \citep{Zamora:2009a}.  It is speculated that the carbon excess
is related to an anomalous helium flash, possibly involving a binary
merger, although models do not yet reproduce this
behavior. \citet{Zamora:2009a} show that late C-R stars are chemically
indistinguishable from C-N stars, and so are also AGB stars.  C-J
stars show enhanced $^{12}$C and Li abundances and may also be a
heterogeneous group; their nature remains unclear \citep{Abia:2000a}.
These classes are likely all {\em intrinsic} carbon stars, responsible
for their own carbon overabundance.  There are also several classes of
{\em extrinsic} carbon stars in which the excess carbon was accreted
from an evolved companion.  These include the barium stars and their
population II counterparts, the C-H stars.  Many of these are found to
have white dwarf companions.

Symbiotic stars are interacting binaries in which a compact object,
usually a white dwarf, accretes from a red giant star
\citep[e.g.][]{Mikolajewska:2007a}.  In most symbiotics, the red giant
appears to be a normal, oxygen-rich star, but of the 188 symbiotics in
the catalog of \citet{Belczynski:2000a} about 6\,\%\ have carbon star
companions.  Half of these are in the Magellanic Clouds, leaving
Galactic carbon star symbiotics quite rare.  One of the defining
characteristics of the symbiotic star population is that they usually
show an emission line spectrum, though exceptions do exist
\citep{Munari:2002a} and may represent a sub-class of low accretion
rate symbiotics.

There is also a small population of symbiotic X-ray binaries (SyXBs)
which instead host a neutron star \citep{Masetti:2006a}.  There are
seven reasonably firm candidates \citep[][and references
therein]{Masetti:2012a}, plus one tentative candidate that is
identified with a carbon star \citep{Masetti:2011a}.  The total
Galactic population of SyXBs has recently been estimated at 100--1000
\citep{Lu:2012a}.  The best studied SyXB is GX~1+4
\citep{Chakrabarty:1997a}, which has a high X-ray luminosity
($\sim10^{37}$\,erg\,s$^{-1}$) and an optical spectrum rich in
emission lines.  As more candidates have been discovered, these
characteristics have been found to be the exception rather than the
rule, and the other candidates have inferred X-ray luminosities of
$10^{32}-10^{34}$\,erg\,s$^{-1}$.  They typically show very hard and
highly variable X-ray spectra.  Presumably because of the low X-ray
luminosity, and absence of a strong UV source to ionize the red giant
wind, the optical spectra are usually lacking in emission lines making
secure confirmation of a SyXB nature challenging.

In addition to these candidate SyXBs, \citet{vandenBerg:2006a}
identified 13 X-ray selected symbiotics in the Bulge.  They found
quite hard X-ray spectra and a paucity of emission lines, suggesting
that this sample may well include some, or even a majority of SyXBs,
or alternatively that X-ray selected white dwarf symbiotics define a
different population from those previously identified optically.  The
objects found by \citet{vandenBerg:2006a} suggest that looking for
X-ray selected cool giants may be an efficient way to expand the
symbiotic parameter space, including identifying more SyXBs, as cool
giants are intrinsically very weak X-ray sources
\citep{Gudel:2004a,Ramstedt:2012a}.

The Galactic Bulge Survey (GBS) is an 0.3--8.0\,keV X-ray survey
performed with {\it Chandra's} ACIS-I camera \citep{Jonker:2011a}. It
was specifically optimized to identify low-luminosity neutron star
X-ray binaries in the Bulge, with a limiting X-ray sensitivity of
$F_{\rm X} > 2.3\times10^{-14}$\,erg\,cm$^{-2}$\,s$^{-1}$ (for a
$\Gamma=2$, $N_{\rm H}=10^{22}$\,cm$^{-2}$ power-law spectrum) which
translates to a luminosity limit of $2\times10^{32}$\,erg\,s$^{-1}$ at
the Bulge distance.  By observing at $|b|>1^{\circ}$, the survey
focuses on regions of reduced absorption relative to that of the
Plane, so we can expect that most of the SyXBs in the survey area
should be detected as X-ray sources.  The modest extinction, coupled
with the high optical and IR brightness of the companion stars in
SyXBs means that we also can expect to detect the optical/IR
counterparts and classify the systems.  There is thus a reasonable
prospect that the GBS can identify most of the SyXBs along its line of
sight and thus provide a near-volume limited survey to test the
population models.

Looking for Bulge symbiotics, SyXBs and white dwarf systems, is also
potentially of interest if objects can be found with carbon star
companions.  Examining the distribution of objects in the Catalog of
Galactic Carbon Stars \citep[CGCS;][]{Alksnis:2001a}, it shows a
pronounced deficit of objects in the vicinity of the Bulge, in spite
of the high density of red giants there.  The entire 12 square degree
field of the GBS contains just 5 objects from the CGCS.
\citet{Blanco:1989a} found only 5 carbon stars among a sample of 2187
late M giants in the Bulge, implying a C-M ratio of just 0.0023.  This
can be compared to the Large and Small Magellanic Cloud C-M ratios of
0.8 and 13.8 respectively \citep{Blanco:1983a}, and a radial gradient
from 0.2 to 0.7 in the disk of M33 \citep{Rowe:2005a}.  Besides their
rarity, the carbon stars found in the Bulge, for example the 34
objects of \citet{Azzopardi:1991a}, are also markedly less luminous
than those in the Magellanic Clouds.  Indeed, \citet{Ng:1997a} went so
far as to suggest that there are no carbon stars in the Bulge, and
that the objects found are actually more distant objects associated
with the Sagittarius Dwarf Spheroidal galaxy seen through the Bulge.
\citet{Whitelock:1999a} has challenged this, however, arguing that the
photometric properties of the Bulge carbon stars are inconsistent with
those of the carbon stars definitely associated with the Sagittarius
Dwarf.  Finding SyXBs associated with Bulge carbon stars can
potentially then shed light on the evolutionary history of these
objects, for example via modeling the spin history of their neutron
stars \citep{Lu:2012a}.

Optical spectroscopy of counterparts to GBS X-ray sources is ongoing
using a number of facilities.  Here we report on the first candidate
SyXB identified by the GBS with spectroscopic classification from
Gemini-South and the Very Large Telescope, CXOGBS~J173620.2--293338
(henceforth CX332 following the source numbering of
\citealt{Jonker:2011a}).  This X-ray source is coincident with a very
red star, 2MASS~J17362020--2933389, also identified as an Optical
Gravitational Lensing Experiment (OGLE) irregular variable (OGLE~IV
BLG\,654.20 36111; \citealt{Udalski:2012a}).  This counterpart shares
the characteristics of Bulge carbon stars.

\section{Data Reduction}

\subsection{CTIO Blanco Mosaic-2 Imaging}

The field of CX332 was observed using the Mosaic-2 camera on the 4\,m
Blanco telescope at the Cerro-Tololo Inter-American Observatory from
2010 July 12 to 18.  We obtained 120\,sec exposures in the SDSS $r'$
filter several times per night as part of an extensive program to
obtain photometry of GBS sources (Britt et al.\ in preparation).  The
images were reduced with the NOAO Mosaic Pipeline \citep{Shaw:2009a}
which added astrometric and (approximate) photometric calibrations
using the USNO B1.0 catalog \citep{Monet:2003a}.

\begin{figure}
\includegraphics[width=3.4in]{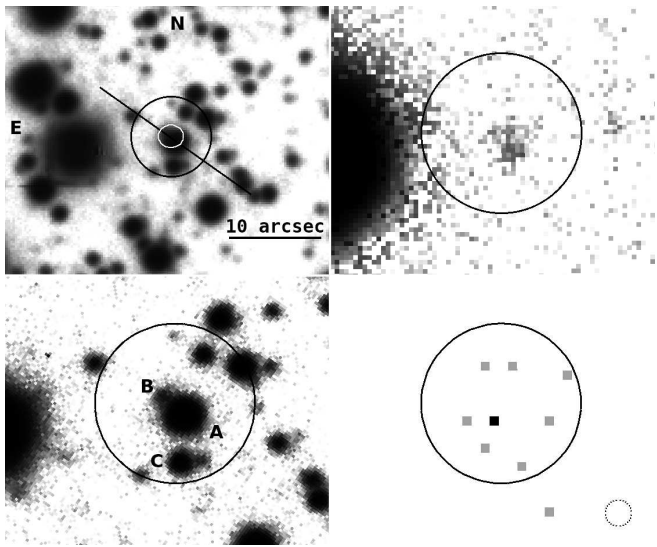}
\caption{Top Left: Average of the best nine Mosaic-2 images.  The
  small white ellipse indicates the 95\,\%\ confidence region
  calculated by combining the {\sc wavdetect} errors from
  \citet{Jonker:2011a} with the {\em Chandra} aspect uncertainty.  The
  larger black circle is the 95\,\%\ confidence region calculated
  according to the prescription of \citet{Hong:2005a}; this is
  reproduced in all the panels of this figure.  The diagonal line
  indicates the alignment of the GMOS slit. Top Right: Closer-up view
  of the Mosaic-2 variance image.  Bottom Left: Close-up of the
  central region from the Gemini-S acquisition image showing the
  second faint star (B) close to the bright carbon star A, as well as
  the southern star C for which a VIMOS spectrum was obtained.  Bottom
  Right: {\em Chandra} ACIS-S 0.3--8.0\,keV image of the same region.
  Grey points are single photon detections, the single black point is
  two photons.  The dashed circle is the 95\,\%\ confidence {\em
    Chandra} aspect uncertainty.}
\label{FinderFigure}
\end{figure}

We combined the nine images with the best seeing to form an average
shown in Fig.~\ref{FinderFigure}.  We checked the astrometry against
ten nearby objects matched with 2MASS objects. The agreement was
excellent, with the Mosaic solution differing from 2MASS by
$(0.01\pm0.02, 0.08\pm0.02)$\,arcsec in RA and Dec.\ respectively.  We
also overlay on Figure~\ref{FinderFigure} the {\em Chandra} position.
Two regions are indicated.  The inner ellipse indicates 95\,\%\
confidence {\sc wavdetect} uncertainties reported by
\citet{Jonker:2011a} combined in quadrature with the {\em Chandra}
aspect uncertainty.  The outer circle is a more conservative 95\,\%\
confidence region based on the prescription of \citet{Hong:2005a}.
The latter explicitly accounts for the degradation of the point-spread
function off-axis, and so may be more realistic for a source observed
7.85\,arcmin off-axis.  However, in Figure~\ref{FinderFigure} we also
show the {\it Chandra} image with individual events localized.  Based
on the event locations, the outer circle actually appears to be an
over-estimate of the plausible uncertainty in the source position.  We
note that while the systematic {\it Chandra} aspect undertainty is
taken account of, it is a negligible contribution to the current
positional uncertainty (see Fig.~\ref{FinderFigure}).  The {\it
  Chandra} observation was too shallow to allow correction of the
aspect uncertainty by cross-correlation with other sources in the
field.

The X-ray position is very close to a bright star (A) and a faint
companion (B) to the north-east; the X-ray localization is not good
enough to discriminate between these.  We measure a brightness for
star A of $r'=17.4\pm0.5$, with the uncertainty dominated by that of
the USNO-based calibration.  Several other stars cannot securely be
ruled out at 95\,\%\ confidence, but are less likely counterparts
since the source is most likely to be near the center of a 95\,\%\
confidence region.  We label the next brightest credible candidate
star C.

We also examined our Mosaic-2 images using the image subtraction
technique \citep{Alard:1998a,Alard:2000a}.  Some $r'$ variability from
star A is detected in the variance images (Figure~\ref{FinderFigure}),
but this is actually quite low amplitude and the variability of star A
is much better sampled by OGLE observations
\citep[][Section~\ref{LightcurveSection}]{Udalski:2012a}.  No signal
is seen in the variance image from stars B or C.

\subsection{Archival Photometry}

Archival photometry is available for star A from a number of sources,
principally in the IR.  We make most use of the Two Micron All Sky
Survey (2MASS) All-Sky Catalog \citep{Cutri:2003a}, corroborated by
the Deep Near Infrared Survey of the Southern Sky (DENIS)
\citep{DENIS:2005a} in the near-IR.  In the mid-IR we have Wide-Field
Infrared Survey Explorer (WISE) All Sky Data Release
\citep{Cutri:2012a} and the Galactic Legacy Infrared Mid-Plane Survey
Extraordinaire (GLIMPSE) Source Catalog \citep{SSC:2009a}.  In
addition, Star A was included in the OGLE Galactic Bulge area as
BLG\,654.20, and a lightcurve is presented by \citet{Udalski:2012a}.
We include their $I$ band magnitude in our photometry database, and
discuss their lightcurve in Section~\ref{LightcurveSection}.  We
summarize the available archival photometry in
Table~\ref{PhotometryTable}

\begin{deluxetable}{lll}
\tabletypesize{\scriptsize}
\tablecaption{\bf Photometry of star A
\label{PhotometryTable}}
\tablewidth{0pt}
\tablehead{\colhead{Survey}  & \colhead{Filter}  & \colhead{Magnitude}}
\startdata
Mosaic-2 & $r'$ & $17.4\phn\phn\pm0.5$ \\
\noalign{\smallskip}
OGLE & $I$         & 15.156 \\
\noalign{\smallskip}
2MASS & $J$        & $11.542\pm0.047$ \\
      & $H$        & $10.130\pm0.046$ \\
      & $K_{\rm s}$ & \phn$9.649\pm0.048$ \\
\noalign{\smallskip}
DENIS & $i$   & $15.062\pm0.09$ \\
      & $J$        & $11.398\pm0.07$ \\
      & $K_{\rm s}$ & \phn$9.483\pm0.05$ \\
\noalign{\smallskip}
GLIMPSE & [3.6] & \phn$9.127\pm0.041$ \\
        & [4.5] & \phn$9.191\pm0.045$ \\
        & [5.8] & \phn$9.067\pm0.025$ \\
        & [8.0] & \phn$8.979\pm0.024$ \\
\noalign{\smallskip}
WISE    & W1 & \phn$8.828\pm0.030$ \\
        & W2 & \phn$8.846\pm0.028$ \\
        & W3 & \phn$8.614\pm0.058$ \\
        & W4 & \phn$7.224\pm0.171$ \\
\enddata
\end{deluxetable}

\subsection{Gemini-S/GMOS Spectroscopy}

Stars A and B were observed with the Gemini Multi-Object Spectrograph
(GMOS) on the Gemini-South telescope as part of a spectroscopic survey
of GBS counterparts.  The acquisition image had a better image quality
than the Mosaic-2 images and clearly resolves stars A and B, so we
also show it in Figure~\ref{FinderFigure}.

Two 450\,s exposures were obtained on 2010 March 18 using the R150
grating spanning the full accessible CCD spectrum at 17\,\AA\
resolution, spread over two CCDs.  The slit was aligned to pass
through both stars A and B (see Figure~\ref{FinderFigure}.  The
spectra were independently reduced using the Gemini {\sc iraf}
package, and using a manual reduction in {\sc iraf}\footnote{IRAF is
  distributed by the National Optical Astronomy Observatories, which
  are operated by the Association of Universities for Research in
  Astronomy, Inc., under cooperative agreement with the National
  Science Foundation.}.  We retained the manual optimal extraction
which was found to be somewhat cleaner.  Flat-fielding used a single
flat taken immediately after the object frames and wavelength
calibration was performed relative to a daytime CuAr arc spectrum
following standard GMOS procedures.

\begin{figure}
\includegraphics[width=3.6in]{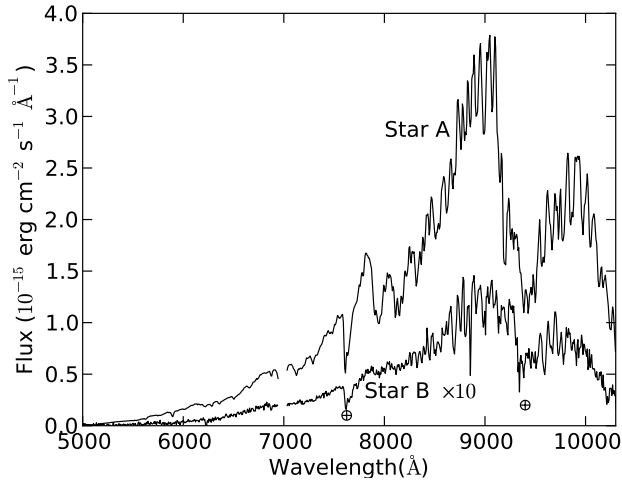}
\caption{Gemini spectra of star A (the carbon star) and B (the fainter
  star).  The flux calibration is crude and primarily intended to
  remove the instrumental response.  The most prominent feature in the
  bright star is the strong CN band at 9100\,\AA, clearly absent in
  the fainter star, which only shows the slightly redder atmospheric
  absorption feature.}
\label{SpecFigOne}
\end{figure}

Extraction of the spectrum of star A proceeded normally using optimal
extraction methods within {\sc iraf}.  The extraction region was
chosen to exclude that contaminated by star B.  To aid comparison with
atlas spectra, we applied a crude flux calibration using the
spectrophotometric standard LTT\,7379 \citep{Hamuy:1994a}.  This was
not observed on the same night, and so does not provide a precise flux
calibration, but suffices to remove most of the effects of the
instrumental response, except at the longest wavelengths.  The full
extracted spectrum combining both CCDs in shown in
Fig.~\ref{SpecFigOne}.  Star A is clearly a carbon star showing strong
features of C$_2$ at 5200\,\AA\ and 5600\,\AA, and multiple CN
features from 5700--6600\,\AA.

Star B was more challenging.  We adapted the methods of
\citet{Hynes:2002a} to extract its spectrum.  We could not use this
approach exactly as there was not a co-aligned, isolated PSF template
star to use.  We began by subtracting a two-dimensional fit to the
sky.  To deal with the blending, we assumed that the profile of the
star A should be spatially symmetric and so reflected the
two-dimensional spectrum around the trace of star A.  This did not
work well in the core of the profile where the interpolation of the
reflected profile was inadequate to describe the data, but did work
well in the wings.  In particular, the profile of star B was well
defined, and isolated from the residuals to the fit in the core of
star A.  We then proceeded to follow \citet{Hynes:2002a} and optimally
extract the spectrum of the star B adding the subtracted light from
the bright star to the effective sky image to ensure proper weighting
of pixels across the profile.  Once extracted, the spectrum was
calibrated as for the star A, and is also shown in
Figure~\ref{SpecFigOne}.  Compared to star A, the spectrum is
relatively featureless, and dominated by the atmospheric A band at
7600\,\AA\ and the water feature at 9300\,\AA.  No emission features
are present.

\subsection{VLT/VIMOS Data}

Stars A and C were also observed with VIMOS \citep{LeFevre:2003a}, an
imager and multi-object spectrograph mounted on the Nasmyth focus of
the 8.2-m European Southern Observatory Unit 3 Very Large Telescope at
Paranal, Chile.  The medium resolution (MR) grism was used to yield a
2.5 \AA/pixel dispersion and a wavelength coverage of $\sim
4800-10000$ \AA.  The use of 1.0 arcsec width slits provided a
spectral resolution of $\sim10$ \AA~FWHM.  The spectroscopic
observations of CX332 were obtained on 2011 July 3 in service mode
under program 085.D-0441(A).  They consisted of two spectroscopic
integrations of 875\,s, along with three flat-field exposures and a
helium-argon lamp exposure for wavelength calibration.  Standard data
reduction was performed with the ESO-VIMOS pipeline \citep{Izzo:2004a}
which averaged the two spectra and automatically extracted the objects
found on the slit.  To handle saturation effects in the spectrum of
CX332, we extracted interactively with the {\sc iraf kpnoslit} package
the reduced 2-D frame that contain both stellar and sky spectra
(pipeline file with product code SSEM). We refer the reader to Torres
et al.\ (in preparation) for further details on the VIMOS spectroscopy
for GBS sources.

\section{Identifying the Optical Counterpart to CX332}

\subsection{Alternative Counterparts}

Star A, the carbon star, is clearly the brightest candidate
counterpart, and is the only optical variable detected within the {\em
  Chandra} error circle, but star B is also close to the center of the
error circle.  As noted above, star B shows an unremarkable spectrum
and no detected variability.  No emission lines are seen, in
particular at the location of H$\alpha$ region.  It is also quite red,
comparable to the star A.  These characteristics suggest a reddened
star in the Bulge, possibly a K-type giant, as no molecular bands are
present to signify an M spectral type.

To estimate the brightness of star B we extracted an average spatial
sky-subtracted profile from the Gemini spectra for the spectral region
7289--8831\,\AA\, approximately corresponding to the $I$ band.  Both
stars show clean, marginally resolved profiles.  We perform a joint
fit to both profiles with two Voigt profiles, with the same Gaussian
and Lorentzian widths and find a brightness ratio of 0.039,
corresponding to a magnitude difference of 3.5 magnitudes.  At the
time of the Gemini observation, the OGLE magnitude of star A was about
15.2, so we estimate that star B is at $I\simeq18.7$.  At the mean
Bulge distance, and reddening $E(B-V)=1.96$ (see
Section~\ref{SEDSection}), this corresponds to $M_I\simeq+1.3$, too
faint to be a giant at 8\,kpc.  It may be a giant on the far side of
the Bulge, a Bulge sub-giant, or a foreground dwarf.  If this were the
X-ray counterpart, with an 8 photon detection ($F_{\rm
  X}\simeq6\times10^{-14}$\,erg\,cm$^{-2}$\,s$^{-1}$;
\citealt{Jonker:2011a}) then it is too X-ray bright for the expected
RS~CVn distribution (Fig.~3 of \citealt{Jonker:2011a}) and lacks the
emission lines expected for a quiescent cataclysmic variable or
intermediate polar.

We also obtained a VLT/VIMOS spectrum of star C with the same
configuration as for star A.  This is also unremarkable and shows no
emission features.  The absorption features detected are typical of G
and early K stars.  Like star B, it is not detected as an optical
variable. We conclude that neither stars B nor C are likely to be the
counterpart to the X-ray source, leaving star A, the carbon star, as
the most probable counterpart.

\subsection{Chance Coincidence Probability}

A complementary approach is to assess the likelihood of a chance
alignment of one of the GBS X-ray sources with a rare object such as a
Bulge carbon star.  The X-ray position reported by
\citet{Jonker:2011a} is $\alpha=264.08432\pm0.00024$,
$\delta=-29.56064\pm0.00014$ where the {\em Chandra} aspect
uncertainty of 0.6\,arcsec (90\,\%\ confidence) has not been included.
The best position for the carbon star is from 2MASS,
$\alpha=264.084174\pm0.000019$ and $\delta=-29.560827\pm0.000017$.
The X-ray/IR positional offset is therefore
$\Delta\alpha=+0.47\pm0.76$\,arcsec, and
$\Delta\delta=+0.69\pm0.49$\,arcsec.  Adding the 1\,$\sigma$ aspect
uncertainty in quadrature, the 1-d offset is $0.83\pm0.69$\,\arcsec.
The X-ray source is clearly consistent with the carbon star.

We can estimate the probability of a chance coincidence based on the
expected surface density of carbon stars.  \citet{Masetti:2011a}
attempted to estimate this for CGCS\,5926 by assuming a total
population of about 2000 carbon stars in the Galaxy, and an effective
surface density of 0.5 per square degree.  This estimate is based on
known sources within the CGCS and is likely a severe underestimate.
The CGCS contains 5 carbon stars within the 12 square area of the GBS,
close to what \citet{Masetti:2011a} assumed.  While it is true that
carbon stars appear to be exceptionally rare in the Bulge, systematic
surveys have found that there are more than this.  We can make some
more realistic estimates based on other studies of carbon stars in the
Bulge.  First, we can compare with the catalog of
\citet{Azzopardi:1991a}.  34 stars were identified in 9 fields (with
some overlap).  On average they find 4 carbon stars per 55\,arcmin
square field implying a density of 4.8 carbon stars per square degree,
an order of magnitude higher than \citet{Masetti:2011a} assumed.
Alternatively, we can note that \citet{Blanco:1989a} found just 5
carbon stars among 2187 M5 or later type giants surveyed in the Bulge,
corresponding to a C-M ratio of 0.0023.  The Besancon Galactic Model
\citep{Robin:2003a} predicts a density of 7300 M5--9 giants per square
degree in the vicinity of CX332.  With the C-M ratio of
\citet{Blanco:1989a}, we then expect around 17 carbon stars per square
degree.  This is a little higher than found by
\citet{Azzopardi:1991a}, but CX332 is closer to the Plane than their
fields were.  Considering 17 carbon stars per square degree to be the
most optimistic prediction, we then expect a probability of
$4\times10^{-7}$ that a carbon star will be found within 1\,arcsec of
CX332, or $7\times10^{-4}$ of a chance alignment within 1\,arcsec of
any of the 1640 GBS X-ray sources.  This supports our conclusion that
star A is most likely to be the true optical counterpart to CX332, and
from here on this work focuses on the properties of this object.

\section{Spectral Classification}

We begin classification of star A by comparing the 5000--7000\,\AA\
spectrum to standard examples from \citet{Barnbaum:1996a}.  For each
one, we convolve the template spectrum with a 10\,\AA\ Gaussian to
provide a closer match to the instrumental resolution of the VLT
spectrum, and scale the spectrum to match our count levels.
Adjustments in this scaling were made between different wavelengths to
correct for differences in the shape of the two spectra.  The relative
strength of features between the template and our target should thus
be independent of uncertainties in calibration and dereddening of our
spectrum.

The primary diagnostic that can be used to reject many of the atlas
spectra is the strength of the C$_2$ Swan bands at 5200\,\AA\ and
5600\,\AA.  Both the overall strength of the bands, and the relative
strength of the two, distinguish different classifications, although
this primarily is a discriminant of the C$_2$ index.

\begin{figure}
\includegraphics[width=3.6in]{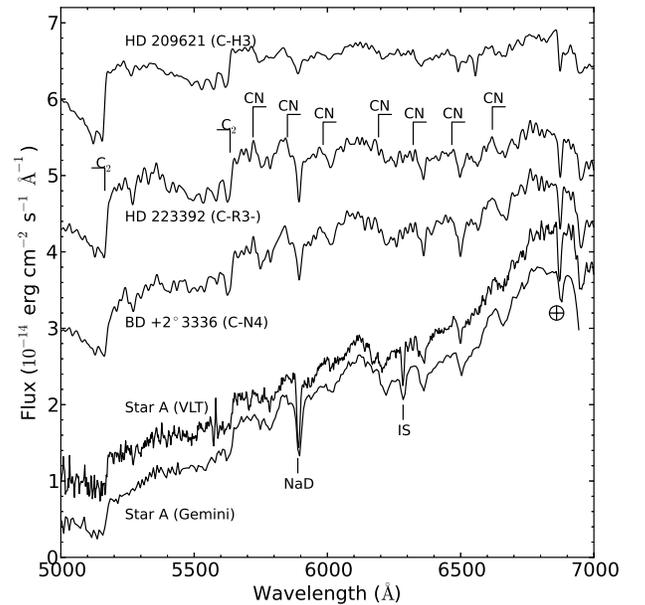}
\caption{Close-up of the red spectra of star A compared to spectra of
  the closest C-H, C-R, and C-N matches from \citet{Barnbaum:1996a}.
  The spectrum of our carbon star has been dereddened.  The spectra of
  the comparison stars were rescaled to approximately the same flux
  and then offset upwards by 2, 3, and 4.5 units.}
\label{SpecFig}
\end{figure}

We find that fair matches are possible with either C-R types or C-N
types.  All of the C-J spectra included by \citet{Barnbaum:1996a} show
much stronger C$_2$ Swan bands (C$_2$ indices of 5 or 5.5) than
observed and can be discounted, as can the barium stars which show
much weaker bands.  The C-H stars are a closer match, but still
notably inferior to C-R or C-N types, at least within the parameter
ranges sampled by \citet{Barnbaum:1996a}.

Among the C-R stars, the best fit was for HD\,223392 (C-R3$-$, C$_2$
4), with a reasonable match also for HD\,76846 (C-R2$+$, C$_2$ 4).
C-R stars with C$_2$ indices of 1.5-3.0 substantially underpredict the
carbon bands, those with C$_2$ index of 5.5 strongly overpredict them.
The best fitting C-N star was BD~$+2^{\circ}$\,3336 (C-N4, C$_2$ 3).
The closest match we could find among the C-H stars was HD\,209621
(C-H3, C$_2$ 4.5).  This reproduces the C$_2$ bands, but underpredicts
the strength of CN.  We show the spectra of star A dereddened with
$E(B-V)=1.96$ (see Section~\ref{DistanceSection}) together with the
best spectral matches in Figure~\ref{SpecFig}.

\begin{figure}
\includegraphics[width=3.6in]{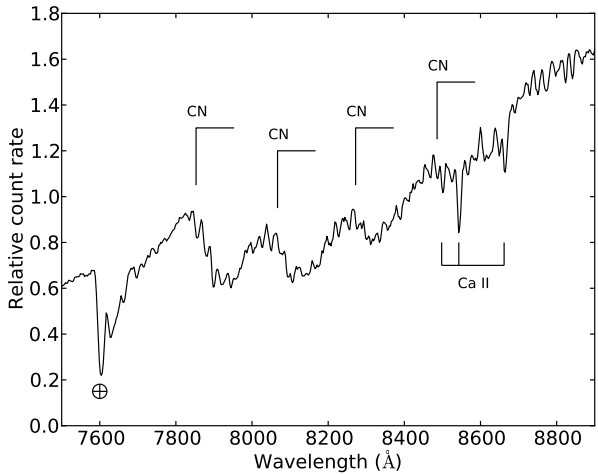}
\caption{Near-IR spectrum of star A obtained with VLT/VIMOS, showing
  the Ca\,{\sc ii} triplet.  We also highlight the CN bandheads
  \citep{Wallace:1962a}.}
\label{NIRSpecFig}
\end{figure}

While we have examined all of the digitized spectrum from
\citet{Barnbaum:1996a}, we are limited in parameter space by the stars
included, and since our primary diagnostic is the strength of the
C$_2$ bands, we mainly are sensitive to the C$_2$ index, rather than
the temperature.  To check this classification we also examined the
VLT $I$ band spectrum in the vicinity of the Ca\,{\sc ii} triplet
which is sensitive to temperature in carbon stars
\citep{Richer:1971a}.  We show this region in Figure~\ref{NIRSpecFig}.
\citet{Richer:1971a} classifies the near-IR spectra into a temperature
sequence from C0--C7.  This system does not exactly correspond to the
system of \citet{Keenan:1993a} which is the basis of the
\citet{Barnbaum:1996a} atlas, but we can identify correspondences
between the two systems where stars are present in both.  In
particular, \citet{Richer:1971a} includes many C5 objects, the
majority of which are also classified as C-N5 by
\citet{Barnbaum:1996a}.

The first thing we note is the presence of multiple prominent CN
bandheads.  This rules out the C0--C2 classifications, in which these
are virtually undetected; the first class in which they are prominent
is C3.  The relative strength of the Ca\,{\sc ii} lines to the CN
bands points to not much later than this, as the 8498\,\AA\ line
should be overwhelmed by CN by C5.  CN and Ca\,{\sc ii} features thus
point to a C3--C4 classification.  The C0--C2 stars identified by
\citet{Richer:1971a} are mostly a mix of C-R and C-H stars with
temperature indices of 1--3.  Since these do not match our spectrum,
we can rule out an early C-R classification, leaving the near-IR
spectra favoring a C-R3--5 or C-N3--5, roughly consistent with the red
classification.  We conclude that star A is a late C-R or early C-N
star.  Both types are considered to be near the bottom of the AGB
\citep{Zamora:2009a}.

\section{Variability}
\label{LightcurveSection}

\begin{figure}
\includegraphics[width=3.6in]{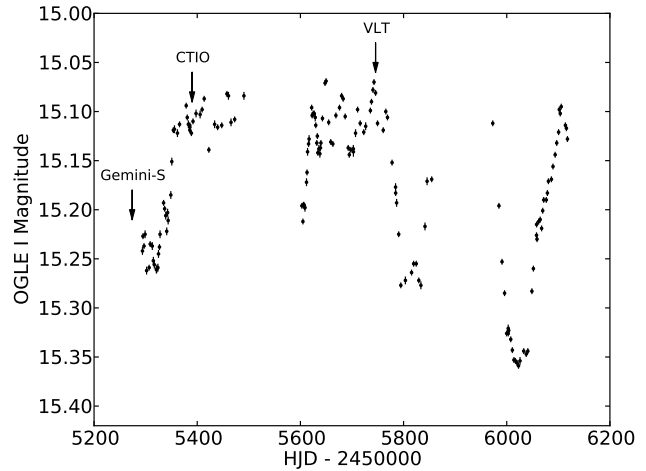}
\caption{OGLE lightcurve of star A, plotted using data provided by the
  OGLE collaboration \citep{Udalski:2012a}.  We also show the times of
  our Gemini-S, CTIO, and VLT observations.  The X-ray observations
  occurred before the period covered by OGLE.}
\label{OGLEFig}
\end{figure}

The optical counterpart to CX332 was identified as an irregular
variable using OGLE data \citep{Udalski:2012a}, with $I$ magnitude
listed as 15.156, and no $V$ detection.  We reproduce the lightcurve
in Figure~\ref{OGLEFig} and indicate the times of our observations.
The behavior seen is quite typical of carbon stars on the AGB, with
the dips indicating periods of dust formation.  This star would be
classified as a slow irregular variable
\citep[Lb;][]{Wallerstein:1998a}.  The dips are relatively shallow
compared to some carbon stars, and so the dust formation is modest and
there is unlikely to be significant local extinction to be accounted
for.  The behavior does indicate that the star is an AGB star, as
variability and dust formation is not seen in the red clump giants
associated with early C-R stars \citep[e.g.\ the tabulation of
][]{Barnbaum:1996a}.  The variability also is additional evidence
against identifying star A as a barium or C-H star in which the carbon
overabundance is a result of binary evolution; these are typically
non-variable.  We note that a few of the later C-H type stars listed
by \citet{Barnbaum:1996a} are irregular or semi-regular variables.
C-H stars appear to be population II counterparts to R stars
\citep[see e.g.][]{Barnbaum:1996a} and so it is likely that they share
the dichotomy between early and late C-R stars \citep{Zamora:2009a}
with some late C-H stars not being products of binary evolution.  In
any case, C-H stars provided a poorer match to the spectrum of star A
than C-R or C-N stars.  The variability then supports the
identification of the counterpart as either a late C-R or C-N type AGB
star.  The low amplitude behavior is quite consistent with other late
C-R stars in particular, for example the C-R4 stars listed by
\citet{Barnbaum:1996a} are all semi-regular (SR) or irregular (Lb)
variables.  Of these, the C-R4 carbon star RV~Sct is included in the
All Sky Automated Survey Catalog of Variable Stars
\citep{Pojmanski:2002a}.  It shows irregular variability with an
amplitude around 0.4\,mags in $V$, and timescales similar to star A.

\section{Spectral Energy Distribution}
\label{SEDSection}

\begin{figure}
\includegraphics[width=3.6in]{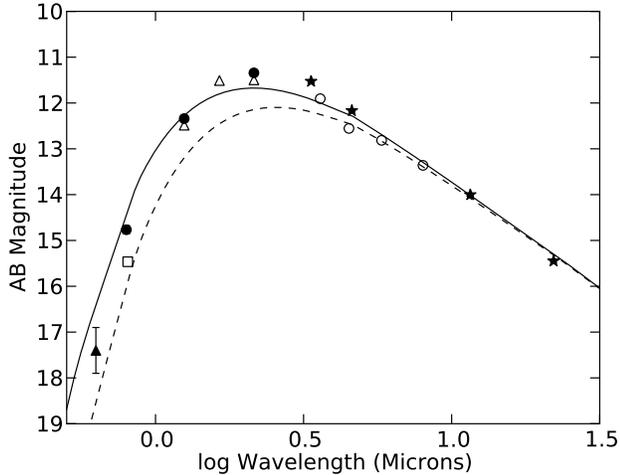}
\caption{Spectral energy distribution of star A.  The filled triangle
  is Mosaic-2 $r'$ data, open square is OGLE $I$, filled circle is
  DENIS $IJK$, open triangle is 2MASS $JHK_{\rm S}$, filled stars are
  WISE, and open circles are from Glimpse.  The dashed and solid lines
  are 3050\,K and 4430\,K blackbody spectra respectively, reddened by
  $E(B-V)=1.96$.  Error bars which are not shown are smaller than the
  size of the points.}
\label{SEDFig}
\end{figure}

We compile the photometry from Table~\ref{PhotometryTable} into a
spectral energy distribution in Fig.~\ref{SEDFig}.  Effective
wavelengths and zeropoints or AB offsets are taken from \citet{Frei:1994a},
\citet{Fukugita:1995a}, \citet{Tokunaga:2005a}, \citet{SSC:2009a}, and
\citet{Jarrett:2011a}.

The line-of-sight extinction to our Bulge fields has been estimated by
\citet{Gonzalez:2011a} and \citet{Gonzalez:2012a} based on red clump
stars in VVV data.  For CX332, we find $E(B-V)=1.96\pm0.28$.
\citet{Wallerstein:1998a} compile effective temperature estimates for
a number of C-N stars, and in particular for a sample of 11 C-N5
stars, the effective temperature is $3050\pm190$\,K.  C-R stars appear
to be systematically hotter.  For three C-R3--5 stars,
\citet{Dominy:1984a} quote $T_{\rm eff}=4430\pm210$\,K.

We therefore overlay blackbody spectra at temperatures of 3050\,K and
4430\,K, both reddened by $E(B-V)=1.96$ using coefficients from
\citet{Cardelli:1989a} in the optical and near-IR and
\citet{Bilir:2011a} for the WISE bandpasses.  These should be
representative of early C-N and late C-R stars respectively.  In as
much as the SED can be crudely characterized by a reddened blackbody
the agreement is fair, so the SED is consistent with a terminally
reddened Bulge late C-R or early C-N star, as inferred
spectroscopically.  The SED follows the blackbodies out to the W4
band (22\,$\mu$m) indicating little if any dust emission is
significant.  This is consistent with the quite modest dust production
implied by the low-amplitude OGLE lightcurve.

\section{Carbon Star Luminosity}
\label{DistanceSection}

We can best estimate the source luminosity using the $K$ band
magnitudes, as these minimize the intrinsic variance in absolute
magnitude among the AGB stars, and also greatly reduce the impact of
extinction uncertainties.  \citet{Wallerstein:1998a} estimated
absolute $K$ magnitudes for a sample of nearby carbon stars using {\it
  Hipparcos} parallaxes.  For the 9 Lb stars, they find $<M_K> = -6.84
\pm 1.18$, a value indistinguishable from the 12 SRb variables in
their sample suggesting that variability type is not a discriminating
factor.  For the 16 stars classified as C-N, the mean is $<M_K> =
-6.95\pm1.13$.

As discussed in Section~\ref{SEDSection}, we adopt
$E(B-V)=1.96\pm0.28$, and hence $A_K=0.67\pm0.10$ for a
\citet{Cardelli:1989a} extinction curve.  The spectral energy
distribution supports this reddening, and in turn a location in the
Bulge.  If we then assume a distance of 8\,kpc, we derive an absolute
magnitude around $M_K=-5.5$.  Allowing for a range in Bulge distances
from 5--13\,kpc results in an absolute magnitude range from $-4.5$ to
$-6.2$.  These values are towards the bottom of the range of the
sample of \citet{Wallerstein:1998a}.  At 8\,kpc it would be more
luminous than the carbon stars SZ~Lep and RU~Pup, so there is no
actual inconsistency.  The observed $K$ band brightness is thus
consistent with a fairly low luminosity AGB type carbon star in the
Bulge.

\begin{figure}
\includegraphics[width=3.6in]{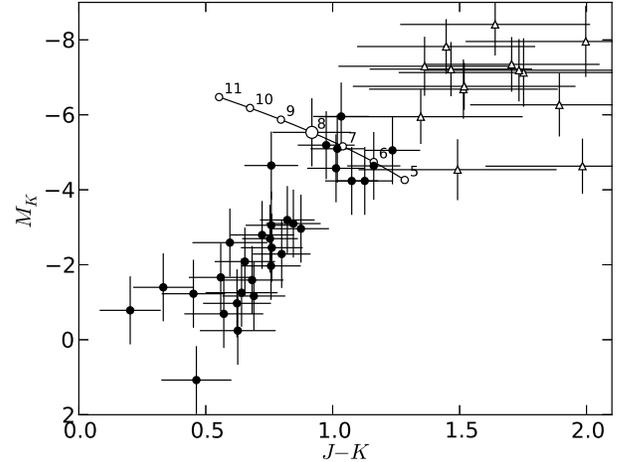}
\caption{IR color-magnitude diagram showing CX332 (open circle)
  compared to the \citet{Azzopardi:1991a} sample of Bulge carbon stars
  (filled circles), and nearby C-N stars from
  \citet{Wallerstein:1998a} with measured parallax distances (open
  triangles).  The dashed line indicates the effect of varying the
  assumed distance of CX332 assuming a linear variation of extinction
  with distance; see text for details.  Annotations correspond to
  1\,kpc steps from 5\,kpc to 11\,kpc.}
\label{AzzopardiFig}
\end{figure}

The inferred low luminosity is not unique to star A, but is a common
characteristic of Bulge carbon stars \citep{Azzopardi:1991a}.  We
examine this systematically in Figure~\ref{AzzopardiFig}.  We assume a
distance of 8\,kpc and extinction values from \citet{Gonzalez:2012a}
to deduce unreddened $J-K$ colors and absolute $K$ band magnitudes of
both star A and Azzopardi's sample.  For comparison, we also combine
absolute magnitudes from \citet{Wallerstein:1998a} with 2MASS colors
to add nearby C-N stars to the diagram.  For the latter, the 2MASS
photometry is saturated, and quoted magnitudes are deduced from
fitting the wings of the profiles.  This introduces larger
uncertainties, but should be useful to crudely indicate the typical
colors.  Since these objects all lie within 1\,kpc, we have assumed
$E(J-K)$ is negligible; for a typical local extinction of
$A_V=2$\,mags per kpc, we then expect $E(J-K)\la0.3$, which is smaller
than the uncertainty in the 2MASS color.  These objects are not
intended to define a complete sample of Galactic carbon stars; as
noted by \citet{Wallerstein:1998a} there are a variety of systematic
selection effects in the {\it Hipparcos} carbon star sample.  Rather
the intent is to indicate where the `classic' C-N type AGB stars lie.

We see quite a striking separation in the diagram, with the Bulge
carbon stars being systematically bluer than the local disk C-N stars,
and a little less luminous.  The equivalent comparison with carbon
stars in the Sagittarius Dwarf galaxy was made in Figure~4 of
\citet{Whitelock:1999a}, where the difference in properties was used
to argue against associating the Bulge carbon stars with that galaxy.
Star A falls securely within the region occupied by the Azzopardi's
Bulge carbon stars, and does not overlap in color with the
distribution of local C-N stars.  This star therefore appears to be a
bona fide member of the Bulge carbon star population.

Among the objects from \citet{Azzopardi:1991a}, we note a division
into two groups in the diagram, with star A lying among the more
luminous objects.  It is possible that the lower group represents the
top of the Bulge carbon star red giant branch, with the upper group
being the AGB.  This is consistent with the
classification of star A as a low luminosity AGB star, and the
evidence for episodes of dust formation in its OGLE lightcurve.

We show the effect of assuming a distance range of 5--11\,kpc in
Figure~\ref{AzzopardiFig} with a dashed line, assuming that extinction
varies linearly with distance.  The latter is a crude assumption, but
it should be remembered that at the latitude of CX332, the line of
sight is still only $\sim200$\,pc above the Plane at the distance of
the Bulge, so has not completely left the disk extinction.  Allowing a
closer distance $\sim7.0$\,kpc would move star A into the middle of
the upper clump of Bulge carbon stars. It may thus be on the near side
of the Bulge.

\section{X-ray Characteristics}

\begin{figure}
\includegraphics[width=3.6in]{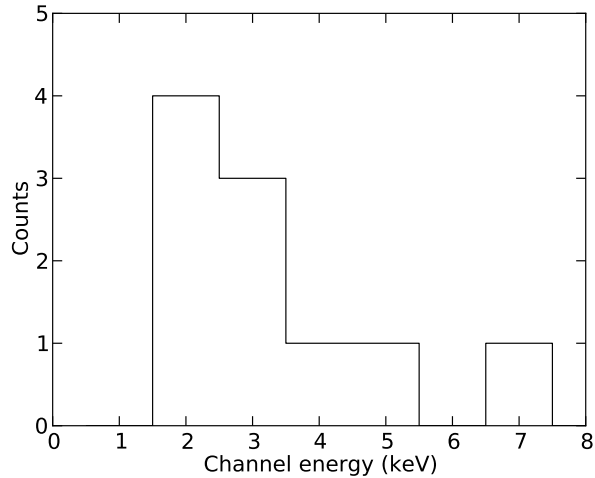}
\caption{Channel energy distribution of {\em Chandra} events in  the
  0.3--8.0\,keV range.}
\label{XSpecFig}
\end{figure}

CX332 was only detected once by the GBS, in observation ID 8693, with
8 reported photons \citep{Jonker:2011a}.  This is clearly insufficient
for a rigorous spectral analysis, but we may still hope to recover
some information about the hardness of the spectrum from the channel
energies of individual photons.  In order to ensure uniformity,
\citet{Jonker:2011a} truncated observations longer than 2\,ks to a
2\,ks length.  Observation 8693 had a livetime of 2.16\,ks, and CX332
was 7.85\,arcmin off-axis, so we re-extract events from the full
exposure time with an aperture radius of 8.1\,arcsec.  We recover 10
events below 8\,keV from the source region and expect about 1 from the
background.  We present a histogram of event channel energies in
Fig.~\ref{XSpecFig}.  The spectrum is clearly hard, with no events
detected below 1.5\,keV, and multiple events above 3.0\,keV.
 
\citet{Muerset:1997a} surveyed {\em ROSAT} spectra of symbiotics.  They
classified them into three groups, $\alpha, \beta$, and $\gamma$.
\citet{Luna:2013a} reexamined the classification scheme in the light
of the harder coverage offered by a {\em Swift} survey.  They added a
fourth $\delta$ class and also identified a number of sources
exhibiting both $\beta$ and $\delta$ components.

Class $\alpha$ shows supersoft emission with negligible counts above
0.4\,keV.  This is believed to originate in quasi-steady nuclear
burning on a white dwarf.  Class $\beta$ shows harder emission
characterized by a thermal spectrum around $10^7$\,K, with most of the
photons at energies below 2.4\,keV.  This may originate in the
interaction of winds from the red giant and the white dwarf's
accretion disk.  The $\gamma$ class of \citet{Muerset:1997a}
originally contained just two objects, the SyXB GX\,1+4 and Hen\,1591
which is suggested to also host a neutron star.  \citet{Luna:2013a}
generalize this to neutron star systems with hard spectra
characterized by Comptonized emission with no emission lines.  The new
$\delta$ class also have hard, absorbed spectra, but with a thermal
spectrum with strong emission lines.  This is attributed to accretion
powered boundary layer emission from a white dwarf surface.

CX332 is clearly not consistent with the $\alpha$ class, and also has
a harder spectrum than is typical of $\beta$ class sources.  Our
observed median event energy is around 2.8\,keV, with only one event
detected with a channel energy below 2.0\,keV.  It most likely fits
into either the $\gamma$ classification, as a neutron star SyXB, or
the $\delta$ class as a white dwarf symbiotic lacking optical emission lines.

If we characterize the spectrum of CX332 with a power-law, then we
find that a $\Gamma = 1.5$ power-law subject to interstellar
extinction does produce the observed median event channel energy.  At
the Galactic center distance, this would imply a luminosity
$\sim9\times10^{32}$\,erg\,s$^{-1}$.  This would be quite typical of
an SyXB, but is higher than the $10^{31}-10^{32}$\,erg\,s$^{-1}$
typically seen in $\delta$ components from white dwarf symbiotics
\citet{Luna:2013a}.

A simple argument can test whether the $L_X \sim
9\times10^{32}$\,ergs\,s$^{-1}$ for CX332 is too high for the $\delta$
class of symbiotic stars.  \citet{Kenyon:1984a} performed detailed
simulations of the optical spectra of symbiotics with hot stars of
different temperatures and luminosities, and found that an accretion
rate of $10^{-9}$\,M$_{\odot}$\,yr$^{-1}$ onto a 1\,M$_{\odot}$ white
dwarf was the lower limit for detection of emission lines in the
optical spectrum, converting to a bolometric luminosity of
$\sim8\times10^{33}$ ergs/s.  \citet{Muerset:1991a} calculate the
bolometric luminosities of a large number of known symbiotic stars,
and show that all but one have a luminosity of the hot component above
10\,L$_{\odot}$, corresponding to $>4\times10^{34}$\,ergs\,s$^{-1}$.
Thus, if the observed X-ray emission is optically thin radiation from
an accretion flow onto a white dwarf (as seen in low-accretion-rate
cataclysmic variables \citep[e.g.][]{Patterson:1985a}, then the total
accretion luminosity can be consistent with the lack of observed
optical emission lines and this possibility cannot be rejected.

\section{Discussion}

We have identified star A, apparently a late C-R type carbon star near
the base of the AGB as the most likely optical counterpart to CX332.
There have been very limited X-ray detections of AGB stars.
\citet{Kastner:2004a} found Mira at a luminosity of
$5\times10^{29}$\,erg\,s$^{-1}$.  \citet{Ramstedt:2012a} examined
observations covering 13 AGB stars and found only two reasonably
confident detections at likely luminosities $<10^{32}$\,erg\,s$^{-1}$.
All three of these AGB stars have quite soft X-ray spectra peaking
around 1\,keV, and in all three cases a binary companion could be the
origin of the X-rays. 
The inferred X-ray luminosity and hardness of CX332 thus appear
inconsistent with intrinsic emission from AGB stars and instead
point to a symbiotic nature.  This allows the possibility of either
a white dwarf or neutron star companion, based on other known
symbiotics.  A black hole companion is also possible in principle,
although would be unprecedented.

The X-ray hardness and luminosity of CX332, and the
lack of optical emission lines of star A, are all very typical of the majority
of known neutron star SyXBs, and CX332 would fit well within this
class.  Currently only one SyXB has been proposed with a carbon star
companion, CGCS\,5926 \citep{Masetti:2011a}, and it is not in the Bulge.

White dwarf symbiotics are typically characterized by soft X-ray
spectra and optical emission lines.  {\em Chandra} and {\em Swift}
observations, however \citep[e.g.][]{Luna:2013a} are showing that some
symbiotics do also have strong, hard components in their spectra, and
some {\em only} have hard components.  In a few cases, e.g.\ NQ~Gem, 
these also lack emission lines in some observations
\citep{Munari:2002a}, resulting in properties quite similar to CX332,
although $\delta$ sources without at least H$\alpha$ emission are
still the exception rather than the rule.  This may be a consequence of
lower accretion rates onto the white dwarf leading to optically
thinner harder X-ray spectra and lower UV luminosities with which to
ionize the red giant wind.  In searching for X-ray sources within the
Bulge, {\em Chandra} will more effectively identify objects such as
these than it will identify classic white dwarf symbiotics with softer X-ray
spectra.

Hence we can expect that an X-ray selected symbiotic population will
have different demographics than one selected optically.  This is
reflected in the sample of symbiotic candidates identified by
\citet{vandenBerg:2006a}, which is dominated by sources with hard
X-ray spectra and a deficit of optical emission lines.  Disentangling
the two populations of SyXBs and X-ray selected white dwarf symbiotics
is then a challenging task observationally, but an essential one if we
are to reliably determine the SyXB population of the Galaxy and test
models such as that of \citet{Lu:2012a}.

\section{Conclusions}

We have examined the closest optical counterpart candidates to the
Galactic Bulge Survey (GBS) source CX332.  The source lies close to an
infrared-bright star that we identify as an AGB-type carbon star, most
likely a late C-R star, or possibly early C-N.  Carbon stars are
extremely rare in the Bulge and so we estimate a probability of only
0.1\,\%\ of finding even one chance coincidence with a carbon star in
the whole GBS.  If the X-ray source is associated with the carbon star
then we have argued that the X-ray luminosity and hardness, and lack
of optical emission lines are more consistent with a symbiotic X-ray
binary (SyXB) rather than with a white dwarf system, although the
latter cannot be confidently ruled out.  If this interpretation is
correct it will be only the second proposed SyXB with a carbon star
companion, and the first based on a secure X-ray detection.

\acknowledgments

This work was supported by the National Science Foundation under Grant
No. AST-0908789.  R.I.H. would like to thank Geoff Clayton for much
guidance in navigating the unfamiliar waters of carbon stars and for
comments on the manuscript and Lauren Gossen for assistance with
observations at Cerro Tololo.

This work is partly based on observations obtained at the Gemini
Observatory, which is operated by the Association of Universities for
Research in Astronomy, Inc., under a cooperative agreement with the
NSF on behalf of the Gemini partnership: the National Science
Foundation (United States), the Science and Technology Facilities
Council (United Kingdom), the National Research Council (Canada),
CONICYT (Chile), the Australian Research Council (Australia),
Ministerio da Ciencia e Tecnologia (Brazil), and Ministerio de
Ciencia, Tecnologia e Innovacion Productiva (Argentina). Our Gemini
Program ID is GS-2010A- Q-61. This work is also based on observations
made with ESO Telescopes at the La Silla Paranal Observatory under
programme ID 085.D-0441(A).

This publication makes use of data products from the Two Micron All
Sky Survey, which is a joint project of the University of
Massachusetts and the Infrared Processing and Analysis
Center/California Institute of Technology, funded by the National
Aeronautics and Space Administration and the National Science
Foundation.  This publication also makes use of data products from the
Wide-field Infrared Survey Explorer, which is a joint project of the
University of California, Los Angeles, and the Jet Propulsion
Laboratory/California Institute of Technology, funded by the National
Aeronautics and Space Administration.  We are grateful to the OGLE
collaboration for making digital lightcurves of candidate counterparts
to GBS sources available.

This research has made use of the SIMBAD database,
operated at CDS, Strasbourg, France, and NASA's Astrophysics Data
System. 


{\it Facilities:} \facility{Blanco}, \facility{CTIO:2MASS},
\facility{CXO}, \facility{Gemini:South}, \facility{OGLE},
\facility{Spitzer}, \facility{VLT:Melipal}, \facility{WISE}.

\end{document}